\documentclass{elsart}

\usepackage{epsfig}

\usepackage{amssymb}

\begin{document}

\begin{frontmatter}



\title{Mechanism of finite-amplitude double-component 
convection due to different boundary conditions}


\author{N.~Tsitverblit\thanksref{*}}

\address{Department of Fluid Mechanics and Heat Transfer, 
Tel-Aviv University, Ramat-Aviv 69978, Israel}
\thanks[*]{Address for correspondence: 
1 Yanosh Korchak Street, apt. 6, Netanya 42495, Israel; 
e-mail: naftali@eng.tau.ac.il}

\begin{abstract}
A new mechanism of double-component
convection is discovered. It emerges in a horizontal layer of
Boussinesq fluid as a stable stratification due to flux boundary
conditions is added to an unstable gradient specified by fixed
boundary values. Driven by this mechanism, steady finite-amplitude
flows reminiscent of \mbox{Rayleigh}---\mbox{Benard} convection
arise even when the background density stratification is stable.
\end{abstract}

\end{frontmatter}



\section{Introduction}
Double-component convection is relevant to as diverse
fields as small-scale oceanography \cite{r:scm}, astrophysics
\cite{r:spghp}, geodynamo \cite{r:bus}, crystal growth \cite{r:crsk},
colloidal suspensions \cite{r:mceg}, and soap films \cite{r:mw}.
Convective flows are also commonly used for testing the ideas
related to transition to turbulence and nonlinear pattern formation
\cite{r:br,r:chbpa}. In addition, double-component flows where a
distinction between the components comes from component different boundary
conditions are of basic significance for large-scale environmental
phenomena. These phenomena range from \mbox{Langmuir} circulations
\cite{r:l83} to the global ocean thermohaline circulation
and climate change \cite{r:stmw,r:dm}.

One major aspect of double-component flows with
different boundary conditions is that the effect of
such conditions can be conceptually analogous to that
of different diffusivities in conventional double-diffusive
convection \cite{r:stnf,r:ver,r:stnl}. In particular, unequal 
diffusion gradients forming in perturbed state due to different 
boundary conditions trigger convection analogously to the 
classical double-diffusion. Such analogy has been 
introduced in \cite{r:twh,r:tbc,r:tlh,r:trp} as
a result of the generalization of an earlier 
idea highlighted in \cite{r:wel}. 

This work reports the existence of a novel mechanism of
double-component convection. The nature of this mechanism stems
from different boundary conditions but is not underlain by the
differential diffusion caused by unequal perturbation gradients of the 
components (differential gradient diffusion). This mechanism has been
identified in a horizontal layer of pure, Boussinesq fluid where
an unstable gradient of the component whose boundary values 
are fixed is combined with a stable stratification due to 
the flux boundary conditions for another component. It 
manifests itself in the form of finite-amplitude steady 
flows before the onset of the respective linear instability.
Not resulting from differential diffusion, such flows arise
even when the net background stratification is stable.
An appropriate perturbation could thus trigger
convection in a broad range of parameters
where such convection could not have
been previously anticipated.

\section{The problem formulation and solution procedures}
Let the diffusivities of the components be equal 
and let a stable stratification due to the component
with flux boundary conditions be combined with an 
unstable gradient of the component with the fixed-value 
conditions. The diffusivities are set equal to eliminate
the effects of the classical double-diffusion, and thus
to examine the effects of different boundary conditions
separately. This is analogous to the approach used in
previous studies of conventional double-diffusive convection.
In most such studies, the components with unequal diffusivities
have not been distinguished from each other in terms of boundary
conditions. Equal diffusivities could, besides, characterize two
solutes, as claimed in \cite{r:pmsns}. The ratio of the viscosity
to diffusivity (which is the \mbox{Prandtl} number, $Pr$) would
then be different from that for temperature in water ($Pr=6.7$) 
used in this work. This parameter, however, is not expected to
have a qualitative effect on the novel phenomenon reported
herein. Equal diffusivities could also be viewed as eddy
coefficients \cite{r:l83,r:dm}, when the ratio between
the viscosity and diffusivity is close to
its value used below.

The background gradients are represented by the \mbox{Rayleigh} 
numbers $Ra=g\alpha|\Delta \overline{T}|d^3/\kappa\nu$ 
and $Ra^{s}=g\beta|\partial \overline{S}/
\partial\overline{x}|d^4/\kappa\nu\equiv\mu Ra$.
Here, $\overline{x}$ is the (dimensional) vertical 
coordinate, $d$ is the width of the horizontal slot, 
$\Delta\overline{T}$ is the (dimensional) difference 
between the values of temperature (standing for the
component with fixed-value boundary conditions) at the lower and upper boundaries, 
$\partial \overline{S}/\partial\overline{x}$ is the 
boundaries-prescribed (dimensional) vertical derivative 
of solute concentration (standing for the component with flux boundary conditions),
$\alpha$ is the coefficient of thermal expansion, 
$\beta$ is the coefficient of the density variation 
due to the variation of solute concentration, $g$ is
the gravitational acceleration, $\nu$ is the kinematic
viscosity, and $\kappa=\kappa_{T}=\kappa_{S}$ is the 
diffusivity of both components. The bar means that 
the respective variable is dimensional. 

\begin{figure}
\centerline{\psfig{file=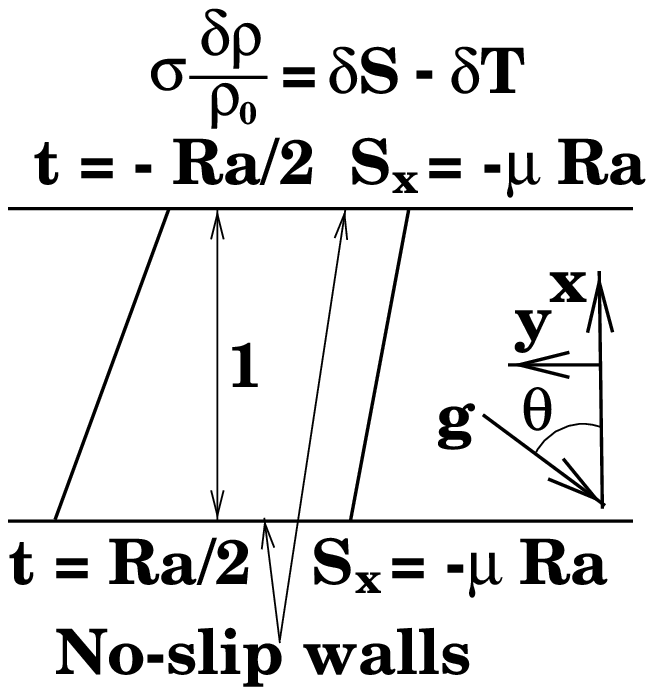,width=6cm}}
\vspace*{1.5cm}
\caption{The problem in a horizontal ($\theta=0$) and 
inclined ($\theta>0$) slot. $\delta\rho=\rho-\rho_{0}$ 
is the variation of the (dimensionless) density, $\rho$, 
due to the variations $\delta S$ and $\delta T$ of solute 
concentration $S$ and temperature $T=(T_1+T_2)/2+t$ with 
respect to their reference values, at which the density 
is $\rho_0$; $T_1$ and $T_2$ are the boundary temperatures,
$\sigma=gd^3/\kappa\nu$. $Pr\equiv\nu/\kappa=6.7$, 
$Le\equiv\kappa_{T}/\kappa_{S}=1$; $\kappa_{T}$ 
and $\kappa_{S}$ ($=\kappa$) are the component 
diffusivities. The fluid is of 
the Boussinesq type.} 
\label{f:g}
\end{figure}

The configuration just described is illustrated
in Fig. \ref{f:g} as $\theta=0$, $\theta$($>0$ in Fig.
\ref{f:g}) is the angle between the direction opposite to
the gravity and that of the across-slot coordinate axis. With
$\Delta_{L}\equiv\partial^{2}/\partial x^{2}+\partial^{2}/\partial y^{2}$,
$\partial_{\tau}\equiv\partial/\partial\tau$,
$\partial_{x}\equiv\partial/\partial x$,
and $\partial_{y}\equiv\partial/\partial y$, the equations describing
the two-dimensional problem in Fig. \ref{f:g} in terms of
streamfunction $\psi$, $t$, and $S$ are:
\begin{displaymath}
Pr(\partial_{\tau}+
\partial_{x}\psi\partial_{y}
-\partial_{y}\psi\partial_{x})\Delta_{L}\psi=
\end{displaymath}
\begin{equation}
(\partial_{x}t-\partial_{x} S)sin\theta-    
(\partial_{y} t-\partial_{y} S)cos\theta+       
Pr\Delta_{L}^{2}\psi,       \label{eq:ns1}
\end{equation}
\begin{equation}
Pr(\partial_{\tau}+\partial_{x}\psi\partial_{y}-
\partial_{y}\psi\partial_{x})\xi_{i}=
\Delta_{L}\xi_{i},\hspace{1cm} i=1,2.        \label{eq:dt}
\end{equation}
Here $\xi_{1}$ and $\xi_{2}$ stand for $t$ and $S$, the
across-slot velocity $u=-\partial_{y}\psi$, the
along-slot velocity $v=\partial_{x}\psi$, and
$\tau$ is the time. This problem was studied
for $\theta=0$ and the periodic boundary conditions with
period $\lambda=\overline{\lambda}/d=2$ in the along-slot
direction by continuation \cite{r:kel} of (finite-difference)
steady solutions in $Ra$ and $\mu$, the numerical approach
was the same as in \cite{r:twh,r:tlh,r:trp}. For clarification
of one relevant issue arising for $\theta>0$ (this case will be
explicitly identified below), continuation in $\theta$ was also used.
$S=0$ was set at the middle points of the across-slot boundaries,
along with the otherwise periodic conditions. Such a condition is 
needed to specify the scale of $S$ and the phase of a nontrivial
steady solution. The time evolution of a linear perturbation
initially imposed on the steady flows was computed to 
examine the solution stability. The stability of the 
conduction base flow to steady disturbances was also 
analyzed for different wave numbers $k=2\pi/\lambda$. 
This was done by searching for the smallest-$Ra$ 
singularity of the matrix resulting from the
application of boundary conditions to the
general solution of the steady, marginal
linear stability problem.

\section{Background}
\begin{figure}
\centerline{\psfig{file=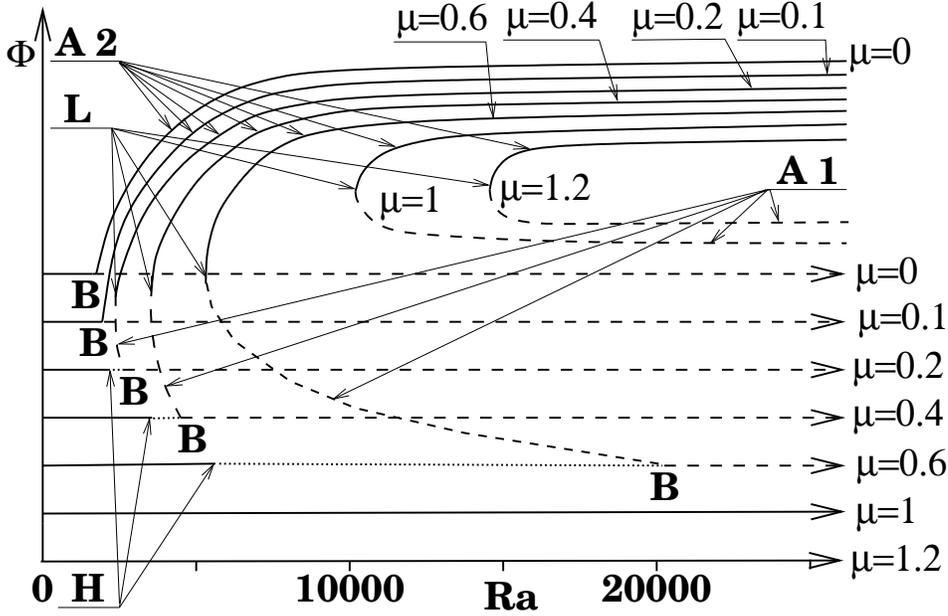,width=12.5cm}}
\vspace*{1.5cm}
\caption{$\theta=0$. Schematic structures of the steady
flows with minimal period $\lambda=2$ for $\mu\in[0,1.2]$
($Pr=6.7$, $Le=1$). $\Phi$ is an abstract measure of the steady
flows that distinguishes between different steady solutions,
specifies the location of the singularities (limit points and
bifurcation points), and represents the flows arising from a
symmetry-breaking bifurcation as a single branch. The background
states are depicted by the horizontal lines with arrows (for $\mu=1.2$,
this is the coordinate axis). The solid lines stand for the stable
solutions. The dashed lines represent the flows being unstable to
either steady or both steady and oscillatory disturbances. The dotted
lines stand for the solutions being unstable to oscillatory disturbances
alone. [Instability of the background state to steady disturbances with 
wave number $k=2\pi$ ($\lambda=1$), arising before that with $k=\pi$ 
($\lambda=2$) at $\mu=0.6$, is not shown.] $B$ is the bifurcation 
standing for the steady linear stability boundary for wave number 
$k=\pi$ ($\lambda=2$); it changes its criticality at $\mu$ just 
below $0.2$ and moves to infinite $Ra$ as $\mu\rightarrow 1$. 
$L$ is the limit point standing for the finite-amplitude steady
stability boundary for the flows with minimal period $\lambda=2$.
It moves to $Ra\approx 25772$ as $\mu$ increases to $1.5$. $A1$
and $A2$ are the unstable and stable (to steady disturbances)
branches associated with the limit point, respectively. 
H is a \mbox{Hopf} bifurcation. For $\mu=1$, 
it arises at $Ra\approx29000$.}
\label{f:bd}
\end{figure}

An infinitesimal disturbance imposed on the
conduction state (Fig. \ref{f:g}, $\theta=0$)
would lead to the formation of unequal perturbation
gradients. Due to the differential gradient diffusion
in perturbed state, a rising (sinking) fluid element
would experience the buoyancy force directed downwards
(upwards) \cite{r:wel}. The buoyancy force is thus expected
to act against the sense of rotation of a small-amplitude
perturbation cell. This permits amplitude growth of the
perturbation cells changing their sense of rotation with 
adequate frequency, as illustrated in \cite{r:wel} for the 
inviscid fluid. (As highlighted in \cite{r:twh,r:tbc}, this
effect makes the present configuration, as well as the configuration
in \cite{r:wel}, analogous to the diffusive regime of conventional
double-diffusive convection \cite{r:stnf,r:ver}.) Such 
oscillatory instability also arises in the viscous fluid 
(Fig. \ref{f:bd}), when the stable flux stratification 
increases. A detailed discussion of the effect of 
viscosity on manifestation of the oscillatory 
instability on different scales is beyond
the scope of this work. Its main idea
is given in \cite{r:tegs}.

(In the framework of current discussion, an oscillatory perturbation 
could be viewed as a standing wave, i.e., as the convective cells 
whose sense of rotation changes periodically in time. The prescribed
across-slot-boundary values of $S$ can prevent traveling-wave
disturbances from being detected in the present formulation. However, 
traveling waves are also expected to arise from such \mbox{Hopf} 
bifurcations as H in Fig. \ref{f:bd} if the translation symmetry of 
the conduction state is allowed for \cite{r:ck}. Their presence,
in particular, may affect the stability of steady branches 
in Fig. \ref{f:bd}.) 

\begin{figure}
\centerline{\psfig{file=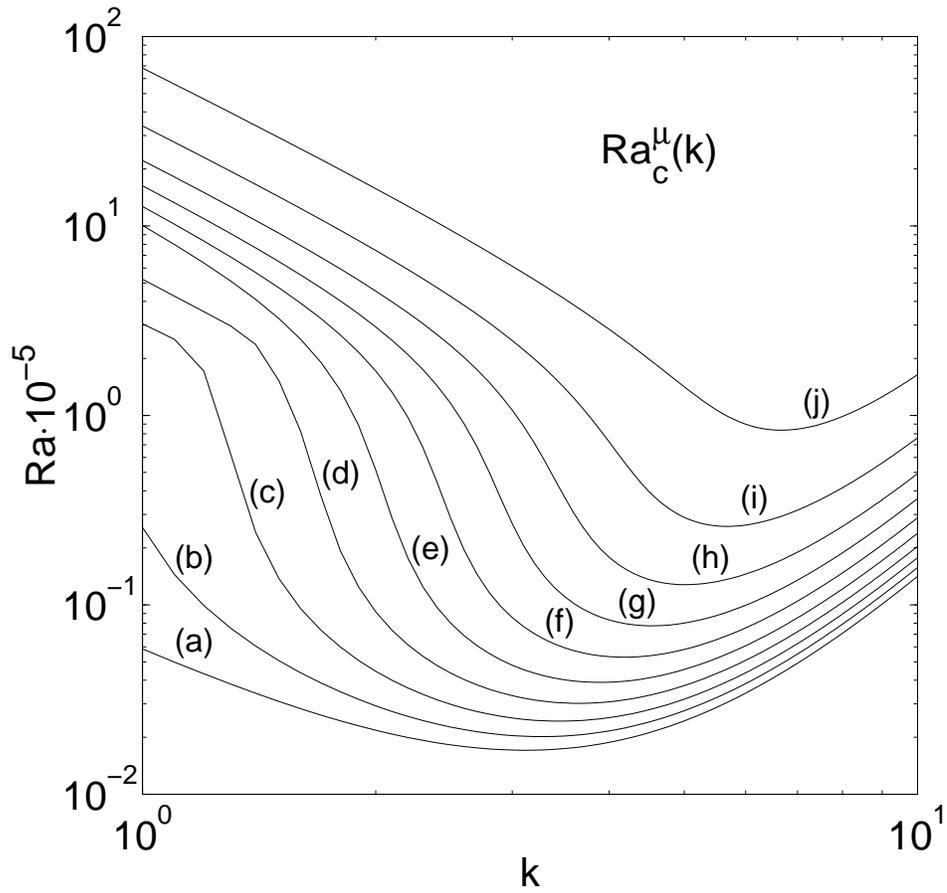,width=12.5cm}}
\vspace*{1.5cm}
\caption{$\theta=0$. Curves of marginal linear stability to steady 
disturbances for different $\mu$ ($Le=1$), $Ra_{c}^{\mu}(k)$. (a) 
$\mu=0$; (b) $\mu=0.1$; (c) $\mu=0.2$; (d) $\mu=0.3$; (e) 
$\mu=0.4$; (f) $\mu=0.5$; (g) $\mu=0.6$; (h) $\mu=0.7$; 
(i) $\mu=0.8$; (j) $\mu=0.9$.}
\label{f:msc}
\end{figure}

Since the differential (gradient) diffusion in the perturbed 
state (Fig. \ref{f:g}, $\theta=0$) results in the growth 
of oscillatory infinitesimal disturbances, it is expected 
to oppose growth of the small stationary perturbations.
(The sense of rotation of a stationary-perturbation cell 
does not change. The cell amplitude could thus grow only
against the above effect of differential gradient diffusion.)
Indeed, $Ra_{c}^{\mu}(k)$ in Fig. 
\ref{f:msc} increasingly exceeds $Ra_{c}^{0}(k)/(1-\mu)$
as $\mu$ grows from $0$. [$Ra_{c}^{0}(k)/(1-\mu)$ are 
the values $Ra_{c}^{\mu}(k)$ would take on if both boundary 
conditions were of the fixed-value type.] The deviation of
$Ra_{c}^{\mu}(k)$ in Fig. \ref{f:msc} from $Ra_{c}^{0}(k)/(1-\mu)$
is particularly pronounced for the large scales (small $k$), 
where diffusion is most effective. With respect to 
infinitesimal steady disturbances, therefore, flux 
boundary conditions for the solute stratification
stabilize the conduction state compared to 
the single-component problem with the
unstable fixed-value gradient.

\section{Results and discussion}
The basic result of this work is that a novel physical
mechanism due to different boundary conditions (Fig. \ref{f:g},
$\theta=0$) gives rise to finite-amplitude convective steady
flows where the conduction state is stable to the infinitesimal
steady disturbances (Fig. \ref{f:bd}). As in convection
resulting from differential gradient diffusion, one element
of this mechanism is disparate responses of the component 
stratifications to convective motion. In the present 
mechanism, however, the feedback to convective
perturbation arises from finite-amplitude
\mbox{Rayleigh}---\mbox{Benard} convection.
This is essentially different from differential gradient
diffusion in \cite{r:twh,r:tbc,r:tlh,r:trp,r:wel}. With
such new feedback, different boundary conditions 
are found to result in a purely nonlinear
manifestation of convection being due
to the statically stable net 
vertical stratification.

\begin{figure}
\centerline{\psfig{file=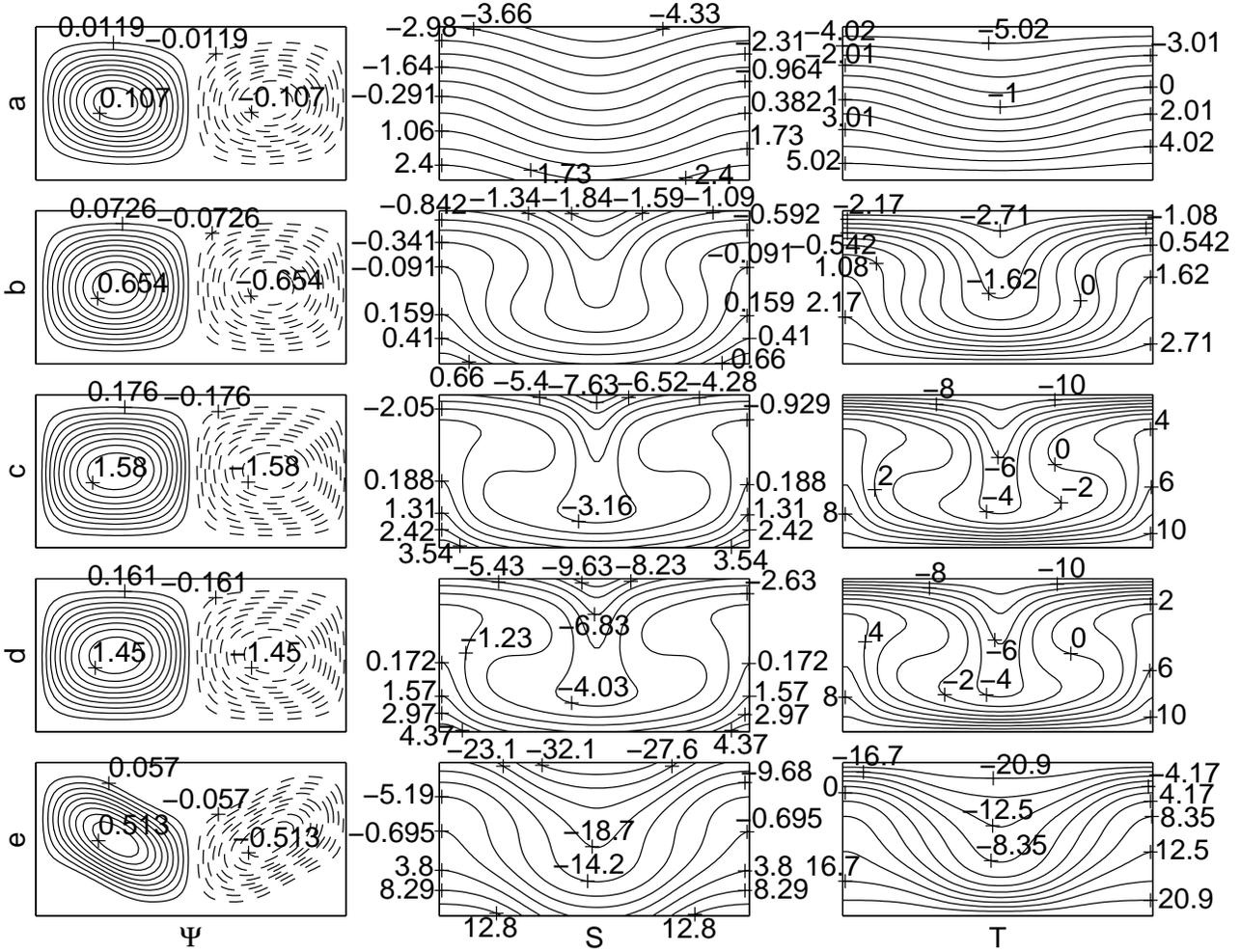,width=17cm}}
\vspace{1.5cm}
\caption{$\theta=0$. Convective steady flows ($Pr=6.7$, 
$Le=1$); $\lambda=2$. $\Psi$: streamlines; $S$: isolines
of solute concentration; $T$: isotherms ($t$ in 
Fig. \ref{f:g}). The solid and dashed streamlines 
represent the clockwise and counterclockwise 
rotation, respectively. The actual values of 
$S$ and $t$ are equal
to $10^3$ times the respective values in the figure.
(a) $\mu=0.6$, $Ra=12046$, branch $A1$ (directly 
unstable); (b) $\mu=0.6$, $Ra=6500$, branch $A2$; 
(c) $\mu=1$, $Ra=24000$, branch $A2$; (d) 
$\mu=1.2$, $Ra=24000$, branch $A2$; (e) 
$\mu=1.2$, $Ra=50090$, branch $A1$ 
(directly unstable).}
\label{f:pst5}
\end{figure}

Finite-amplitude convective flows are illustrated in 
Fig. \ref{f:pst5}. As the convection amplitude increases 
[Fig. \ref{f:pst5}(a),(b)], the ratio of the across-slot 
solute concentration scale to such scale in the background 
state decreases. This is the result of an increasing number 
of solute isolines moving ''outside'' the flow domain, 
especially in the regions of across-slot motion. Such behavior 
is associated with the flux conditions permitting solute 
isolines to cross the boundaries. The respective ratio for
the temperature, however, remains equal to one, even when
convection becomes well-developed [Fig. \ref{f:pst5}(c),(d)]. 
The fixed-value boundary conditions maintain the vertical 
temperature scale by preventing the intersection of isotherms 
with the boundaries. They also increase the thermal gradient 
near the wall towards which the across-slot component 
of convection is directed.

The unstable density gradients thus formed in the regions of across-slot motion
in Fig. \ref{f:pst5} substantially exceed $Ra_{c}^{0}(\pi)$ in Fig. \ref{f:msc}(a).
[$Ra_{c}^{0}(k)$ in Fig. \ref{f:msc}(a) represents the onset of \mbox{Rayleigh}---\mbox{Benard}
convection.] Such gradients result in a horizontal density difference between two streamline
points, as in \mbox{Rayleigh}---\mbox{Benard} convection. This gives rise to (positive)
convective feedback that maintains the disparity between component gradients. Such
feedback forms even when the net background stratification is neutral or stable
[Fig. \ref{f:pst5}(c),(d)]. In such cases of the present formulation, linear
steady instability does not arise, as in the scenario first proposed in
\cite{r:rd}.  As $\mu$ increases, the cells at the smaller-amplitude 
branch, $A1$, change their form to utilize the regions 
with maximal gradient disparity more efficiently 
[Fig. \ref{f:pst5}(e)].

The analogy in the physics of oscillatory instability between 
the present configuration and the diffusive regime of conventional 
double-diffusive convection \cite{r:ver} does not seem to apply to 
finite-amplitude steady instability in both these problems. 
The finite-amplitude mechanism in \cite{r:ver} hinges on the 
disparity between component diffusivities. Such a disparity makes
the unstable temperature gradient relatively insensitive to a 
convective perturbation. This gives rise to the feedback maintaining 
convection. In the present mechanism, differential gradient diffusion
plays only a stabilizing role as the (feedback) unstable density 
stratification arises from the interaction of perturbation with
boundary conditions. In contrast to \cite{r:ver}, in particular,
subcritical steady convection arises in the present problem
even when the density stratification is statically stable.

Nonlinear \mbox{Rayleigh}---\mbox{Benard} convection also 
gives rise to feedback in the finite-amplitude steady instability
in binary fluid \cite{r:blks}, if the separation ratio is 
negative. However, the binary-fluid finite-amplitude mechanism 
is underlain by the dependent nature of the stabilizing background
(Soret) solute gradient, rather than by boundary conditions.
Such solute gradient is thus largely destroyed when its 
conduction-state relation to the unstable 
temperature gradient is relaxed by a 
finite convective perturbation.

Let $\theta>0$ in Fig. \ref{f:g} and let $\mu=1$. When $\theta=\pi/2$,
finite-amplitude steady flows arise due to differential gradient diffusion
\cite{r:trp}. For large enough convection amplitudes ($Ra\geq\sim 20000$),
the dissimilarity between the nature of such flows at $\theta=\pi/2$ and that
of finite-amplitude steady convection at $\theta=0$ was found to give rise to
a region of hysteresis in $\theta$. Within this region, different convective steady
flows with $\lambda=2$ coexist. One of these solutions was continued from $\theta=\pi/2$
in decreasing $\theta$. For say $Ra=31000$, it extends to $\theta\approx0.38\pi/2$.
Another such flow was continued from $\theta=0$ in increasing $\theta$. For $Ra=31000$,
it extends to $\theta\approx0.51\pi/2$. For the same $Ra$($\approx31000$), such
hysteresis phenomenon was not found between $\theta=\pi/2$ and $\theta=\pi$,
where steady convection is driven only by differential gradient diffusion
\cite{r:twh,r:tbc,r:trp}. A detailed analysis of the flow
transformations will be reported separately.

\section{Conclusions}
Underlying a fundamentally new convective mechanism, different 
boundary conditions thus result in finite-amplitude steady convection 
arising in a horizontal layer of pure fluid well before the respective 
linear instability. Such nonlinear convective flows also exist when the
vertical density stratification is statically stable. The new mechanism 
has to remain basically relevant for $Le\neq 1$, as well as under different
$Pr$ (as say for two solutes) and stress-free boundaries, among other changes.
It also raises the issue of three-dimensional effects. Another issue is the
existence of an analogous effect (as well as of the effects of differential
gradient diffusion) when the buoyancy forces are replaced by the forces due
to surface tension. All this leads to a new perspective on the role of
convection and different boundary conditions in double-component
fluid systems, including the large-scale systems relevant
to global environmental processes.

\newpage
\newpage
\newpage
\newpage
\newpage

\end{document}